\documentclass[]{article}

\usepackage{defaults}

\begin{document}

\author{Manuel Koller \\ Seminar for Statistics, ETH Zurich, Zurich, Switzerland}
\title{Nonsingular subsampling for S-estimators with categorical predictors}
\maketitle

\begin{abstract}	
  An integral part of many algorithms for S-estimators of linear regression
  is random subsampling. For problems with only continuous predictors
  simple random subsampling is a reliable method to generate initial
  coefficient estimates that can then be further refined. For data with
  categorical predictors, however, random subsampling often does not work,
  thus limiting the use of an otherwise fine estimator. This also makes the
  choice of estimator for robust linear regression dependent on the type of
  predictors, which is an unnecessary nuisance in practice. For data with
  categorical predictors random subsampling often generates singular
  subsamples. Since these subsamples cannot be used to calculate
  coefficient estimates, they have to be discarded. This makes random
  subsampling slow, especially if some levels of categorical predictors
  have low frequency, and renders the algorithms infeasible for such
  problems. This paper introduces an improved subsampling algorithm that
  only generates nonsingular subsamples. We call it \emph{nonsingular
    subsampling}. For data with continuous variables it is as fast as
  simple random subsampling but much faster for data with categorical
  predictors. This is achieved by using a modified LU decomposition
  algorithm that combines the generation of a sample and the solving of the
  least squares problem.
\end{abstract}

\section{Introduction}
In a nutshell, a random subsampling based algorithm for S-estimates of
linear regression does the following. It takes a random sample of the
observations of size equal to the number of predictors $p$, resulting in a
square design matrix, solves a ``least squares'' problem on this reduced
data set (which gives $0$ residuals in this case) and refines the resulting
parameter estimate using a redescending M-estimate of regression with a
simultaneous scale on the whole data set. This is repeated for a
pre-specified number of times. This final S-estimate is then taken to be
the one that resulted in the smallest scale estimate. Definitions of these
methods are given in Section~\ref{sec:def}.

The random subsampling algorithms described above work well for continuous
data. For categorical data, however, they are often slow or fail
completely.  The problem lies in the generation of \emph{good} subsamples,
i.e., square subsamples that contain no collinearities. Otherwise the
subsample has to be discarded, since it does not generate a well defined
starting value. We illustrate the problem by means of a simple
example. Imagine a simple one-way ANOVA with 3 groups and 3 repetitions
each. Using treatment contrasts to encode the grouping structure, we get
the following least squares problem.
\begin{equation*}
  \left(\begin{array}{c}
      y_1 \\ y_2 \\ y_3 \\ y_4 \\ y_5 \\ y_6 \\ y_7 \\ y_8 \\ y_9
    \end{array}\right)
  = \left(\begin{array}{ccc}
      1 & 0 & 0 \\
      1 & 0 & 0 \\
      1 & 0 & 0 \\
      1 & 1 & 0 \\
      1 & 1 & 0 \\
      1 & 1 & 0 \\
      1 & 0 & 1 \\
      1 & 0 & 1 \\
      1 & 0 & 1 
    \end{array}\right) 
  \left(\begin{array}{c}
      \beta_1 \\ \beta_2 \\ \beta_3
    \end{array}\right) +
  \left(\begin{array}{c}
      \varepsilon_1 \\ \varepsilon_2 \\ \varepsilon_3 \\ \varepsilon_4 \\ 
      \varepsilon_5 \\ \varepsilon_6 \\ \varepsilon_7 \\ \varepsilon_8 \\ 
      \varepsilon_9
    \end{array}\right)
\end{equation*}
In this example, the size of the subsample is $3$. It is clear that the
coefficients $\vec\beta$ can only be estimated if at least one observation
of each group is part of the subsample. From the total number of possible
subsamples, $84$, only $27$ correspond to a nonsingular subsample.
Therefore, even in this very simple example, the probability of discarding
a subsample because of collinearities is about two thirds. This probability
is much higher when some levels of factors have low frequencies. Then an
excessively large number of subsamples is required, rendering the simple
random subsampling algorithms unfeasible for such data sets.

Instead of discarding the whole sample, we propose to solve this problem by
dropping the observation causing the singularity and continue sampling. We
call this refined subsampling strategy \emph{nonsingular subsampling}.

\citet{maronna2000categorical} have proposed another approach on this
problem. They solve it using a combination of M and S-estimates, called
\emph{M-S-estimates}. The categorical part is estimated separately with
M-estimates. The continuous part is estimated using S-estimates. Their
approach works well for data that contains only purely categorical and
continuous variables. For interactions of categorical and continuous
variables, however, it is not clear how to split the data. Using the
M-estimate for the interaction will result in a loss of robustness, while
using the S-estimate will again produce singular subsamples.

In the next section we introduce the notation and provide definitions for
all methods used. Then we develop the basic algorithms for M-estimates and
S-estimates. In Section~\ref{sec:csub} we explain the nonsingular
subsampling algorithm. Finally, we conclude with
Section~\ref{sec:conclusions}.

\section{Notation and definitions}
\label{sec:def}
Consider the standard multiple linear regression model, 
\begin{equation*}
  y_i = \vec x_i\tr\vec \beta + \varepsilon_i, \quad 
  \varepsilon\sim\mathcal{N}\fn{0, \sigma^2}, \quad
  i = 1, \dots, n,
\end{equation*}
with $e_i$ i.i.d., independent of $\vec x_i$ and $\vec\beta$ of length
$p$. Throughout this text we assume that the design matrix $\bld X$,
combining all $\vec x_i$ into one large matrix, is of full rank $p$. We
denote the residuals as $r_i\fns{\vec{\hat\beta}} = y_i - \vec
x_i\tr\vec{\hat\beta}$.

\emph{Simultaneous M-estimates of regression and scale} are defined as the
solutions $\vec{\hat\beta}$ and $\hat\sigma$ to,
\begin{align}
  \label{est:M}
  \sum_{i=1}^n \psi\fn{\frac{r_i\fn{\vec{\hat\beta}}}{\hat\sigma}}\vec x_i
  = {} & \vec 0, \\
  \label{est:M-scale}
  \frac{1}{n}\sum_{i=1}^{n}\chi\fn{\frac{r_i\fn{\vec{\hat\beta}}}{\hat\sigma}} 
    = {} & \kappa,
\end{align}
where $\kappa$ is a tuning constant, $\chi$ is a so-called $\rho$-function,
and $\psi$ is a $\psi$-function. A $\rho$-function, as defined in
\citet{MarRMY06}, is assumed to be a nondecreasing function of $\abs{r}$,
with $\rho\fns{0} = 0$ and strictly increasing for $r > 0$ where
$\rho\fns{r} < \rho\fns{\infty}$. If a $\rho$-function is bounded, we
assume $\rho\fns{\infty}$ = 1 and call it a \emph{redescending}
$\rho$-function. A $\psi$-function is the derivative of a $\rho$-function
and is usually standardized so that $\psi'\fns{0} = 1$. A solution
$\hat\sigma$ to \eqref{est:M-scale} for a given vector $\vec r$,
$\hat\sigma\fns{\vec r}$, is called \emph{M-estimate of scale}. The
solutions to \eqref{est:M} for redescending $\rho$-functions are local
minima of the corresponding optimization problem and are called
\emph{redescending M-estimates of regression}.

\emph{S-estimates of regression} are the parameter value $\vec{\hat\beta}$
that minimizes the M-estimate of scale $\hat\sigma = \hat\sigma\fns{\vec
  r\fns{\vec{\hat\beta}}}$ of the associated residuals,
\begin{equation}
  \label{est:S}
  \vec{\hat\beta} = \argmin_\beta \hat\sigma\fn{\vec r\fn{\vec\beta}}.
\end{equation}
The maximal breakdown point $(1 - p/n)/2$ of the S-estimate is attained
when using $\kappa = (1 - p/n)/2$. Note that solutions of \eqref{est:S} are
always also the solution to a simultaneous M-estimation of regression and
scale problem. For an introduction to M-estimation and details about
S-estimates, we refer to \citet{MarRMY06}.

A \emph{LU decomposition} of a matrix $\bld A$ is defined as the product of
a unit lower-triangular matrix $\bld L$ and an upper-triangular matrix $\bld
U$. Such a decomposition does not always exist, therefore in practice one
usually uses an \emph{LU decomposition with partial pivoting}. This is an
LU decomposition of the matrix $\bld A$ where the rows have been reordered
by a permutation matrix $\bld P\tr$. The ordering is chosen in a way that
minimizes numerical errors. The decomposed matrix can then expressed as,
\begin{equation*}
  \bld A = \bld P\bld L\bld U.
\end{equation*}
This form is useful to solve systems of linear equations $\bld A\vec\beta =
\vec y$. Using the LU decomposition this problem can be solved directly using
a forward and a backward elimination.

\section{Basic algorithms}
\label{sec:algorithms}
First we describe algorithms to compute M-estimates of regression and
M-estimates of scale. The equations \eqref{est:M} and \eqref{est:M-scale}
are preferably solved using iterative reweighting algorithms. As outlined
in \citet{MarRMY06}, one can rewrite \eqref{est:M} and \eqref{est:M-scale}
to take the form of a weighted least squares problem and a weighted sample
variance. The weights then correspond to the respective robustness weights.
Starting from some suitable initial estimate, the iterative reweighting
algorithm then simply repeats the following two steps until
convergence. The first step consists of the computation of the weights
using the results from the previous iteration. In the second step the
weighted problem is solved using the weights computed in the first step.
Convergence of this algorithm is usually not a problem. For redescending
M-estimates the algorithm converges to a local minimum. One can easily get
an algorithm solving the simultaneous M-estimation of regression and scale
problem by combining the two iterations.

We now come to the computation of S-estimates. Combining the above
mentioned properties of S-estimates and M-estimates, we can derive a simple
algorithm that we will later use as the basis for further improvements. See
Algorithm~\ref{alg:basic} for a summary of the rest of the paragraph.
Firstly, note that since any S-estimate is also a solution to
\eqref{est:M}, we can restrict the minimization to parameter vectors
$\vec\beta$ that solve \eqref{est:M} and \eqref{est:M-scale}
simultaneously. Secondly, we use the iterative reweighting algorithm
described above to generate such solutions from a set of initial
estimates. Such a set of initial estimates may be found by solving all
subproblems involving only $p$ observations. Finally, we get the desired
S-estimate as the solution resulting in the smallest scale estimate.

Since for subproblems with $p$ observations the design matrix is square,
the least squares problem reduces to a set of linear equations. It has only
a defined solution if the design matrix is invertible, i.e., there are no
collinearities. We may simply discard all singular subproblems. Note that the
check for singularity does not require an additional step. If the system of
linear equations of the subproblem can be solved we may continue, otherwise
we discard the subsample.

\begin{algorithm}
  \DontPrintSemicolon
  \KwData{$\vec y$, $\bld X$.}
  \KwResult{$\vec{\hat\beta}$, $\hat\sigma$.}
  \nl $\hat\sigma \leftarrow \infty$\;
  \nl \For{all subsets $J \subset \{1,\dots,n\}$ of size $p$}{
    \nl \If{design matrix $\blds{X}{J}$ contains no collinearities}{
      \nl $\vec{\hat\beta}_0 \leftarrow \blds{X}{J}^{-1}\vec y_J$\;
      \nl $\vec{\hat\beta}_J, \hat\sigma_J \leftarrow$ Solve
      \eqref{est:M-scale} and \eqref{est:M} using iterative reweighting
      algorithm starting from $\vec{\hat\beta}_0$.\; 
      \nl \If{$\hat\sigma_J < \hat\sigma$}{
        \nl $\hat\sigma \leftarrow \hat\sigma_J$\;
        \nl $\vec{\hat\beta} \leftarrow \vec{\hat\beta}_J$\;
      }
    }
  }
  \caption{Basic algorithm for the computation of S-estimates.}
  \label{alg:basic}
\end{algorithm}

This algorithm using exhaustive resampling, i.e., running over all possible
subproblems, as shown in Algorithm~\ref{alg:basic}, is of course only
suitable for small problems. For large problems, it is neither feasible nor
sensible to look at all the subsamples. Instead, it is enough to consider
only set of random subsamples. Depending on the expected proportion of
outliers, we can use simple combinatorics to determine how many random
subsamples are required to select at least one outlier-free subsample with
a probability of, say, $0.999$. The number of subsamples grows
exponentially with $p$. Taking $1000$ random subsamples has proven to work
well in practice. Exact numbers are given in Table~5.3 of
\citet{MarRMY06}. The same reference also summarizes many more
optimizations. Worth mentioning is the paper by \citet{salibian2006fast}
where they develop a complete strategy for computing S-estimates, dealing
also with very large data sets.

\section{Nonsingular subsampling}
\label{sec:csub}
Simple random subsampling algorithms have the drawback that they cannot
guarantee the generation of nonsingular subsamples, i.e., without
collinearities. They work on a simple trial-and-error basis. In the
following, we propose an algorithm that produces only nonsingular
subsamples. The algorithm we propose has the advantage that it is much
faster than simple random subsampling algorithms for hard problems, without
sacrificing any time for easy problems.

The nonsingular subsampling algorithm merges the two steps of generating
the random subsample and solving the system of linear equations. The latter
consists of computing a LU decomposition (for a definition, see
Section~\ref{sec:def}) and then solving two triangular linear systems of
equations. Instead of generating the whole subsample at once, we propose to
select observation by observation. A new observation is only added to the
subsample if it is not collinear to the observations already
selected. Proceeding in this way, we will always end up with a nonsingular
subsample. The speed up is achieved by using a modified LU decomposition
algorithm. It computes the LU decomposition observation by observation,
without having to recompute any results of previously selected observations
if one observation needs to be dropped. So if the random subsample is
nonsingular in the first place, e.g., for continuous predictors, the
algorithm does the same as the simple random subsampling algorithms. But
for singular subsamples, we can avoid restarting from scratch, simply
dropping the observation and continue with the next candidate is enough.

The modified LU decomposition algorithm is based on the so-called Gaxpy
variant of the LU decomposition algorithm as found in
\citet{golub1996matrix}. It is of the same complexity as other LU
decomposition algorithms. The Gaxpy variant delays the computation of
columns of $\bld U$ until they are actually needed. To compute the $i$th
column of $\bld L$, we need only columns $1$ to $i$ of $\bld U$. In case a
singularity is detected, it is therefore enough to only repeat the last
step using a new observation / column. Results obtained prior to this step
do not need to be recomputed.

\begin{algorithm}[p]
  \DontPrintSemicolon
  \KwData{$n\times p$ matrix $\bld X$, response vector $\vec y$,
    singularity treshold $\varepsilon$.} 
  \KwResult{Return code ($0$ for success, otherwise number of failing step),
    initial estimate $\vec{\hat\beta}$.}
  \SetKwFunction{Perm}{perm}
  \SetKw{Kwin}{in}
  \SetKw{KwGoto}{Goto}
  \tcR{Initialize variables, pivoting table $\vec p$, selected subsample
    index vector $\vec s$ }\;
  \nl $\bld U \leftarrow \bld 0$; $\bld L \leftarrow \bld I$;
  $\vec p \leftarrow 1:p$; $\vec s \leftarrow 1:p$; $k \leftarrow 1$\;
  \tcR{Permutate observations randomly}\;
  \nl $\vec t \leftarrow$ \Perm{$1:n$}\;
  \nl $\bld A \leftarrow \bld X_{\vec t, 1:p}\tr$\;
  \nl $\vec y \leftarrow \vec y_{\vec t}$\;
  \nl \For(\tcR{Find non-singular subsample and calculate LU
    decomposition}){$j$ \Kwin $1$ \KwTo $p$}{
    \nl \lIf{$j == 1$}{
      $\vec v_{1:p} \leftarrow \bld A_{1:p, k}$\;
    } \label{csub:step}
    \nl \Else{ 
      \tcR{(Forward)solve to get required column of $\bld U$}\;
      \nl $\bld U_{1:j-1,j} \leftarrow \bld L^{-1}_{1:j-1,1:j-1} \bld A_{1:j-1,k}$\;
      \nl $\vec v_{j:p} \leftarrow \bld A_{j:p,k} - \bld L_{j:p, 1:j-1}\bld U_{1:j-1,j}$\;
    }
    \nl \If{$j < p$}{
      \tcR{Find pivot}\;
      \nl $\mu \leftarrow \argmax_{l=j}^p |\vec v_{l}|$\;
      \nl \If{$|\vec v_{\mu}| \geq \varepsilon$}{
        \tcR{Subsample is still non-singular}\;
        \nl $\vec p_j \leftarrow \mu$\;
        \nl $\vec s_j \leftarrow k$\;
        \tcR{Swap elements of $\vec v$ and rows of $\bld A$}\;
        \nl $\swap{\vec v_{j}}{\vec v_{\mu}}$\;
        \nl $\swap{\bld A_{j,k+1:n}}{\bld A_{\mu,k+1:n}}$\;
        \tcR{Update $\bld L$}\;
        \nl $\bld L_{j+1:p,j} \leftarrow \vec v_{j+1:p} / \vec v_{j}$\;
        \tcR{Swap rows of $\bld L$}\;
        \nl $\swap{\bld L_{j,1:j-1}}{\bld L_{\mu,1:j-1}}$\;
      }
    }
    \nl \If{$|\vec v_{j}| < \varepsilon$}{
      \tcR{Singularity detected: skip this column and try again if possible}\;
      \nl \If{$k < n$}{
        \nl $k \leftarrow k + 1$\;
        \nl \KwGoto{\ref{csub:step}}\;
      }
      \nl \Else(\tcR{Return with an error}){
        \nl \Return{j}\;
      }
    }
    \nl $\bld U_{j,j} \leftarrow \vec v_{j}$\;
    \nl $k \leftarrow k + 1$\;
  }
  \tcR{Solve $X_{\vec s,1:p}\inv\vec y_{\vec s}$ and undo pivoting}\;
  \nl $\vec{\hat\beta} \leftarrow \bld L\trinv\bld U\trinv\vec y_{\vec s}$\;
  \nl \lFor{$j$ \Kwin $p-2$ \KwTo $0$}{
    $\swap{\vec{\hat\beta}_j}{\vec{\hat\beta}_{\vec p_j}}$\;
  }
  \nl \Return{$0$, $\vec{\hat\beta}$}\;
  \;
  \caption{Constrained subsampling using modified Gaxpy variant of LU
    decomposition. We use vector index notation to select subvectors and
    submatrices. The index vector $1:p-1$ in $\vec v_{1:p-1}$ indicates
    that an operation only acts on the elements $1$ to $p-1$ of the
    vector.}
  \label{alg:csub}
\end{algorithm}

The nonsingular subsampling algorithm is shown in Algorithm~\ref{alg:csub}.

The numerical stability of the algorithm can be improved by using a matrix
preconditioning technique that reduces the condition number of the design
matrix.  We propose to use a method called \emph{equilibration}, described,
e.g., in \citet{demmel1997applied}. Tests have shown that it is enough to
apply this method to the complete design matrix, even if only submatrices
are used later on.

\section{Conclusions}
\label{sec:conclusions}
Current algorithms for S-estimates of linear regression have trouble coping
with categorical predictors, especially if interactions of continuous and
categorical predictors are involved. These issues can be avoided by using
nonsingular subsampling instead of simple random subsampling. By merging
the steps of generating the subsample and fitting the least squares
problem, the new subsampling algorithm can generate nonsingular subsamples
much more efficiently than simple random subsampling for data sets with
categorical predictors. Comparing the runtimes of S-estimates using
nonsingular subsampling and M-S-estimates showed only a modest increase of
computing time (around $10\%$) even for quite large designs ($n = 8088, p =
340$, $2$ of them continuous predictors). The nonsingular subsampling
algorithm is implemented in the \texttt{lmrob} function of the \textsf{R}
package \texttt{robustbase} from version \texttt{0.9-3}.

\clearpage

\bibliographystyle{elsarticle-harv}
\bibliography{constr}

\begin{thebibliography}{5}
\providecommand{\natexlab}[1]{#1}
\providecommand{\url}[1]{{#1}}
\providecommand{\urlprefix}{URL }
\expandafter\ifx\csname urlstyle\endcsname\relax
  \providecommand{\doi}[1]{DOI~\discretionary{}{}{}#1}\else
  \providecommand{\doi}{DOI~\discretionary{}{}{}\begingroup
  \urlstyle{rm}\Url}\fi
\providecommand{\eprint}[2][]{\url{#2}}

\bibitem[{Demmel(1997)}]{demmel1997applied}
Demmel J (1997) Applied Numerical Linear Algebra. Society for Industrial and
  Applied Mathematics

\bibitem[{Golub and {Van Loan}(1996)}]{golub1996matrix}
Golub GH, {Van Loan} CF (1996) Matrix computations, 3rd edn. The Johns Hopkins
  University Press, Baltimore

\bibitem[{Maronna and Yohai(2000)}]{maronna2000categorical}
Maronna RA, Yohai VJ (2000) Robust regression with both continuous and
  categorical predictors. Journal of Statistical Planning and Inference
  89(1–2):197 -- 214, \doi{10.1016/S0378-3758(99)00208-6}

\bibitem[{Maronna et~al(2006)Maronna, Martin, and Yohai}]{MarRMY06}
Maronna RA, Martin RD, Yohai VJ (2006) Robust Statistics, Theory and Methods.
  Wiley, N.Y.

\bibitem[{Salibian-Barrera and Yohai(2006)}]{salibian2006fast}
Salibian-Barrera M, Yohai V (2006) {A fast algorithm for S-regression
  estimates}. Journal of Computational and Graphical Statistics 15(2):414--427

\end{thebibliography}

\end{document}